\documentstyle[prb,aps]{revtex}

\input epsf
\epsfverbosetrue

\begin{document}
\renewcommand{\textfraction}{0.02}

\def\Journal#1#2#3#4{{#1}\ {\bf #2}, #3 (#4)}
\def\CPL{Chem.\ Phys.\ Lett.}
\def\PRL{Phys.\ Rev.\ Lett.}
\def\PRB{Phys.\ Rev.\ B}
\def\PRD{Phys.\ Rev.\ D}
\def\ZPB{Z.\ Phys.\ B}
\def\SSC{Solid State Commun.}
\def\NAT{Nature}
\def\PB{Physica B}
\def\PC{Physica C}
\def\EPJB{Eur.\ Phys.\ J. B}
\def\LCM{Less.-Common Met.}
\def\EPL{Europhys.\ Lett.}
\def\JS{J.\ Super.}
\def\JCG{J.\ Cryst.\ Growth}
\def\ACB{Acta Cryst.\ B}
\def\JAC{J.\ Appl.\ Crystallogr.}
\def\JLTP{J.\ Low Temp.\ Phys.}
\def\JPCM{J.\ Phys.\ Cond.\ Matt.}
\def\JPSJ{J.\ Phys.\ Soc.\ Jpn.}
\def\MRB{Mat.\ Res.\ Bull.}
\def\S{Science}

\twocolumn[\hsize\textwidth\columnwidth\hsize\csname
@twocolumnfalse\endcsname
\title{Hole distribution for (Sr,Ca,Y,La)$_{14}$Cu$_{24}$O$_{41}$ ladder compounds
studied by x-ray absorption spectroscopy}
\author{N. N{\"u}cker, M. Merz, C. A. Kuntscher, S. Gerhold, and S. Schuppler}
\address{Forschungszentrum Karlsruhe, IFP, P. O. Box 3640, D -- 76021 Karlsruhe, Germany}

\author{R. Neudert,$^a$ M. S. Golden, and J. Fink}
\address{Institut f{\"u}r Festk{\"o}rper- und Werkstofforschung Dresden, P. O. Box 270016,
D -- 01171 Dresden, Germany}

\author{D. Schild}
\address{Forschungszentrum Karlsruhe, INE, P. O. Box 3640, D -- 76021 Karlsruhe, Germany}

\author{S. Stadler, V. Chakarian, J. Freeland,$^b$ and Y. U. Idzerda}
\address{Naval Research Laboratory, Code 6345, Washington, DC 20375}

\author{K. Conder}
\address{Laboratorium f{\"u}r Festk{\"o}rperphysik, ETH Z{\"u}rich, CH -- 8093 Z{\"u}rich, 
Switzerland}

\author{M. Uehara, T. Nagata, J. Goto, and J. Akimitsu} 
\address{Department of Physics, Aoyama-Gakuin University, Chitosedai, Setagaya-ku,
Tokyo 157-8572, Japan}

\author{N. Motoyama, H. Eisaki, and S. Uchida}
\address{Department of Superconductivity, The University of Tokyo, Bunkyo-ku, 
Tokyo 113-0033, Japan}

\author{U. Ammerahl}
\address{Laboratoire de Physico-Chimie des Solides, Universit\'{e} de Paris-Sud, F -- 91405 
Orsay CEDEX, France, and\\
II. Physikalisches Institut, Universit{\"a}t zu K{\"o}ln, Z{\"u}lpicher Str.
77, D -- 50937 K{\"o}ln, Germany}

\author{A. Revcolevschi}
\address{Laboratoire de Physico-Chimie des Solides, Universit\'{e} de Paris-Sud, F -- 91405 
Orsay CEDEX, France}

\date{Received August 8, 2000}

\maketitle

\begin{abstract}
The unoccupied electronic structure for the Sr$_{14}$Cu$_{24}$O$_{41}$ 
family of two-leg ladder compounds was investigated for different partial 
substitutions of Sr$^{2+}$ by Ca$^{2+}$, leaving the nominal hole count constant,
and by Y$^{3+}$ or La$^{3+}$, reducing the nominal hole count from its full value of 
6 per formula unit. Using polarization-dependent x-ray 
absorption spectroscopy on single crystals, hole states on both the chain and 
ladder sites could be studied. While for intermediate hole counts all 
holes reside on O sites of the chains, a partial hole occupation on the ladder 
sites in orbitals oriented along the legs is observed for the fully doped compound 
Sr$_{14}$Cu$_{24}$O$_{41}$\@. On substitution of Ca for Sr orbitals within the 
ladder planes but perpendicular to the legs receive some hole occupation as well.
\end{abstract}
]

\section{Introduction}

\begin{figure}[tb]
\noindent
\begin{minipage}[t]{86mm}
\hspace*{14mm}
\vspace*{5mm}
\leavevmode \epsfxsize=60mm \epsfbox{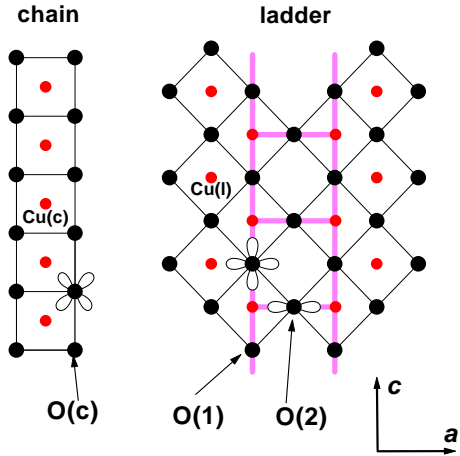}
\caption[]{Structural elements of Sr$_{14}$Cu$_{24}$O$_{41}$\@. Shown are 
chain units consisting of Cu(c) and O(c) sites, and ladder units consisting 
of Cu(l) and O(1,2) sites, with the ``legs'' and ``rungs'' highlighted
for one ladder as shaded lines. The chain and ladder layers are stacked 
alternatingly along the $b$ axis, with Sr sites in-between, to form the 
actual crystal structure.} \label{structure}
\end{minipage} 
\end{figure}

Spin-$1/2$ Heisenberg ladder compounds have been the subject of extensive 
investigations over the last few years,\cite{Matsud} first and foremost  
due to their magnetic peculiarities. These are caused by the magnetic 
interactions within the legs and the rungs which, in principle, may be 
very different for the various sublattice units. Assuming, e.\ g., a coupling 
across the rungs that is much larger than that along the legs leads to the 
formation of a spin gap for compounds with even numbers of legs and to a tendency 
for ``hole pairing''\@. Both features turn out to persist even for fairly similar 
coupling constants as in fact found for many actual ladder compounds.
Early on, superconductivity was predicted 
for even-leg ladders\cite{rice} and was eventually found for 
Sr$_{0.4}$Ca$_{13.6}$Cu$_{24}$O$_{41.8}$.\cite{uehara} 
This compound is a member of the more complex family 
(Sr,Ca)$_{14}$(CuO$_2$)$_{10}$(Cu$_2$O$_3$)$_7$, which contains layers of 
Cu$_2$O$_3$ two-leg ladders extending along the crystallographic $c$ axis
(see Fig.\ \ref{structure}); the ladder layers are stacked along the $b$ axis and 
alternate with layers of edge-sharing CuO$_2$ chains.\cite{carron,siegrist} The 
observation of superconductivity is highly remarkable as this compound is a cuprate 
but does not contain the CuO$_2$ planes so characteristic for high-temperature 
superconductors (HTSCs)\@. It is, therefore, not surprising that these findings 
initiated a flurry of experimental and theoretical investigations on these 
compounds, their electronic structure, and its relationship to the magnetic 
properties.

Due to its composition, Sr$_{14}$Cu$_{24}$O$_{41}$ is doped by 6 holes, 
which are generally believed to reside essentially in the chain. 
Perhaps the simplest indication for this is

\noindent
provided by bond-valence-sum (BVS) 
calculations,\cite{katobvs} giving higher BVS values for holes in the chains. 
Similar to the HTSCs, the holes are 
expected to be mainly of O $2p$ character as the large Hubbard 
$U$ for Cu $3d$ states prevents them from having much Cu character. 
On the experimental side, resistivity and magnetic measurements on
``undoped'' La$_6$Ca$_8$Cu$_{24}$O$_{41}$, with purely a Cu$^{2+}$
valence and thus no holes, as well as on compounds with increasing
Sr content for La and a corresponding number of doped holes led to the 
conclusion that {\it all} doped holes reside on the chains of 
Sr$_{14}$Cu$_{24}$O$_{41}$, giving a Cu$^{2.6+}$ valence 
there.\cite{carter} Neutron scattering,\cite{Matsuda,Regnault} 
however, points to a spin dimerization along the chain, which in
turn was interpreted as a signature for a CDW-like arrangement of one 
hole for every other Cu(c) site of the chain. This means that in total 
only 5 out of the 6 doped holes reside in the (CuO$_2$)$_{10}$ chain subunit 
of Sr$_{14}$Cu$_{24}$O$_{41}$, leaving one hole for the (Cu$_2$O$_3$)$_7$ 
ladder unit. Optical investigations\cite{Uchida,ruzicka} are also consistent with 
a hole count of one in the ladder unit of Sr$_{14}$Cu$_{24}$O$_{41}$; upon Ca 
substitution this was found to increase to up to 2.8 holes. 
A recent Cu NMR study\cite{magishi} relates measured spin correlation lenghts 
to the average distance between holes and determine a hole occupation on
the ladder sites up to 1.75 holes for Sr$_{2.5}$Ca$_{11.5}$Cu$_{24}$O$_{41}$\@. 
The formation of dimers was supported by further NMR\cite{Takigawa} and
NQR\cite{ohsugi} experiments, and the underlying charge ordering was observed
by x-ray diffraction.\cite{Cox}   
Also, NMR measurements\cite{tsuji} suggest that the spin gap of the ladder 
unit decreases continuously over the whole substitutional range from zero 
holes to the fully doped and heavily Ca-substituted compounds until it 
almost reaches zero, while the spin gap of the chain remains essentially 
constant, lending support to the notion that it is indeed the ladders that
carry the
\begin{figure}[tb]
\noindent
\begin{minipage}[t]{86mm}
\leavevmode
\epsfxsize=86mm \epsfbox{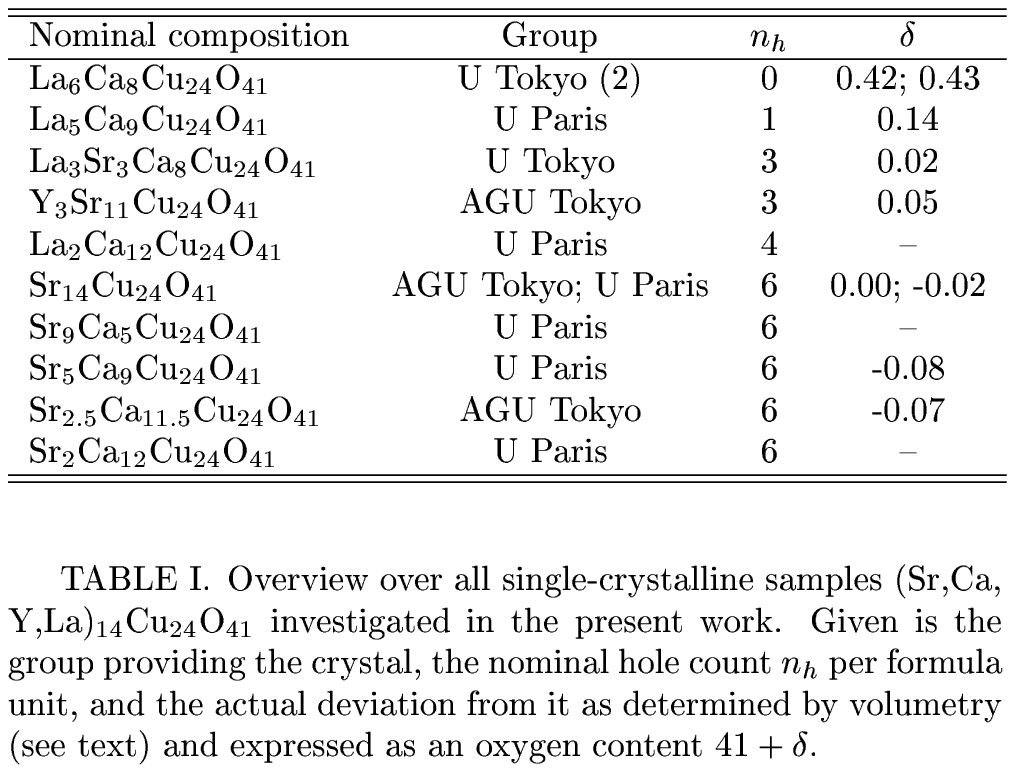}
\end{minipage}
\end{figure}

\noindent
superconductivity. On the other hand, other NMR experiments\cite{magishi}
and neutron scattering\cite{katano5} find a constant spin gap upon Ca 
substitution for the ladders as well. Chemical pressure by Ca substitution 
for Sr as well as physical pressure reduces the interlayer 
distances\cite{Pachot} and increases the twisting of the CuO$_2$ 
chain.\cite{Isobe} As a consequence, about 30 - 50\% 
of the O(c) atoms reduce their distance to 
the nearest Cu(l) atom to a value comparable with the typical 
Cu-O$_{\rm apex}$ distance for HTSCs.\cite{Isobe} A transfer of holes 
from the chains to the ladder sites concurrent with Ca substitution
was predicted theoretically\cite{Theorie} and was also inferred from
various experiments.\cite{Uchida,ruzicka,Experiment} 

Despite all this wealth of information on the ladder compounds, however, 
there remains a considerable amount of controversy, and a more {\em direct}
determination of the site-specific hole distribution is still 
missing. Both the evolution of hole distributions with increasing
number of holes and with increasing Ca content with its structural
implications is important to know. In the present study, 
polarization-dependent x-ray absorption spectroscopy is applied, which is 
a proven method to obtain such information. The results, in turn, may help 
to better understand the electronic structure and possibly the onset of 
the insulator-metal transition as well as the appearance of superconductivity 
in these systems.

\section{Experiment and results}

Single crystalline samples (typically about 60 mm$^3$ in size) were cut from
several cm long crystals grown by the traveling
solvent floating zone (TSFZ) method;\cite{growth,Ammerahl} their compositions are 
shown in Table I\@. The formal valence of 
Cu in the samples was determined by a precision volumetric 
method,\cite{Conder} where an overall accuracy of $< 5 \times 10^{-4}$ can 
be achieved. An average Cu valence +2+$\varepsilon$ determined in this way gives a 
total hole count $n_{\rm Cu}$=24$\varepsilon$, and any deviation from the nominal 
number of holes expected from stoichiometry, $n_h$, is accounted for by 
simply taking the O content, which could not be determined indepen-
\begin{figure}[tb]
\noindent
\begin{minipage}[t]{86mm}
\hspace*{4mm}
\vspace*{3mm}
\leavevmode \epsfxsize=75mm \epsfbox{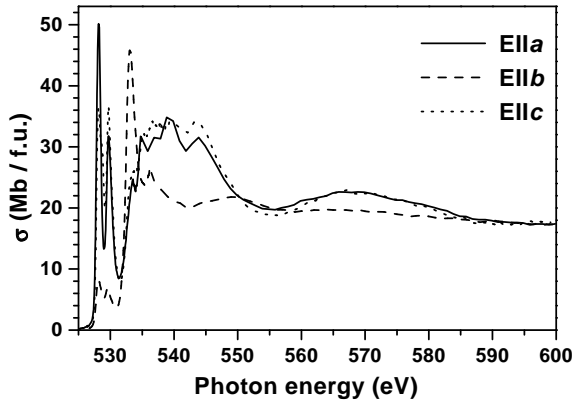}
\caption[]{Polarization-dependent O $1s$ absorption spectra for 
Sr$_{14}$Cu$_{24}$O$_{41}$\@. For {\bf E}$\|$$a,c$ two large pre-peaks
appear below the main absorption edge at about 535 eV; for {\bf E}$\|$$b$ these 
pre-peaks are far smaller and a dominant, Sr-related structure appears at the main 
absorption edge.} \label{fullrange}
\end{minipage} 
\end{figure}

\noindent
dently, as 
41+$\delta$ atoms per formula unit with $\delta$=($n_{\rm Cu}$-$n_h$)/2\@. The 
largest $\delta$ observed in the present study occurred for the 
La$_6$Ca$_8$Cu$_{24}$O$_{41.43}$ samples and corresponds to 0.86 extra holes. An 
investigation by scanning electron microscopy (SEM) and microspot Auger analysis 
revealed that a few percent of La in these samples is segregated 
as La-rich precipitates, illustrating the fact that a high La content leads to 
phase instability in these compounds.\cite{Ammerahl} The $\delta$=0.14 found 
for La$_5$Ca$_9$Cu$_{24}$O$_{41.14}$ may result from the same effect. For none of 
the other compositions traces of such phases could be detected by SEM or 
x-ray diffraction; also, much smaller values for $\delta$ were observed 
(Table I)\@.

Polarization-dependent near-edge x-ray absorption fine structure (NEXAFS) 
measurements were performed at the National Synchrotron Light Source (NSLS) 
at the NRL/NSLS beamline U4B\@. The energy resolution in the range of the
O 1$s$ NEXAFS was set to 210 meV. To ensure a flat and shiny surface,
several slices ($\approx$200 nm each) were cut off from the samples
using the diamond blade of an ultramicrotome, and the samples were then 
transferred to ultrahigh vacuum. The data were recorded at room temperature
and in fluorescence detection mode probing a depth of about 100 nm. The spectra 
were corrected for incident flux variations and for self-absorption effects,\cite{troeger}
and were normalized to tabulated cross sections\cite{yeh} at about
70 eV above the edge where the absorption is atomic-like and structureless.
For energy calibration of the O $1s$ spectra, total electron yield data of 
NiO were taken simultaneously and referred to a NiO standard\cite{eels} from
electron energy-loss spectroscopy with a reproducibility significantly better 
than the experimental energy resolution.

The aim of this experiment is to investigate the hole distribution
in the three inequivalent O sites of
(Sr,Ca,Y,La)$_{14}$Cu$_{24}$O$_{41}$\@. With each of the inequivalent sites, a
different Cu-O bond configuration is associated ({\it cf}.\ Fig.\ \ref{structure}): 
a Cu-O-Cu 
interaction geometry with about 90$^{\circ}$ for the O(c) sites of the chains,
similar to the situation in Li$_2$CuO$_2$;\cite{Sapina,neudert} straight bonds 
(180$^{\circ}$) connecting the Cu(l) sites in the two legs of a single ladder 
{\it via} the rung O(2) sites; and the situation for the O(1) sites of the 
legs, which form bonds to each of the three neighboring Cu(l) atoms.
Assuming that, analogous to the situation in HTSCs, only $\sigma$ bonds formed 
between Cu $3d$ and O $2p$ states contribute spectral weight near the Fermi 
level, $E_F$, there are only five different O $2p$ orbitals in the $a$,$c$ plane
(see Fig.\ \ref{structure}) 
which may contribute spectral weight in O $1s$ NEXAFS\@. The excitation 
process in NEXAFS involves the highly localized O $1s$ core level; moreover, 
dipole selection rules apply. As a consequence, orienting the electric field 
vector of the incident radiation, {\bf E}, parallel to the crystallographic 
axes allows one to investigate {\em specifically} the O $2p$ orbitals oriented 
likewise, and thus gain site-specific information on the hole distribution.

\smallskip

Fig.\ \ref{fullrange} depicts the O $1s$ absorption and its polarization 
dependence over the full measured energy range for Sr$_{14}$Cu$_{24}$O$_{41}$ 
as a representative example for the whole set of compositions. The two large peaks 
observed near threshold for the in-plane, {\bf E}$\|$$a,c$ spectra are due to Cu-O 
hybridization and are strongly suppressed for the out-of-plane, {\bf E}$\|$$b$
spectrum. The latter exhibits right at the main absorption edge a strong feature 
which is most likely associated with Sr-O hybrids connecting the chain and ladder
layers. In the extended (EXAFS) range the variation of the absorption coefficient 
$\sigma$ is very different from that for the in-plane geometry, illustrating that 
the spatial environment of O in the direction perpendicular to the planes is very 
different from that within the planes.

\begin{figure}
\noindent
\begin{minipage}{178mm}
\leavevmode
\epsfxsize=178mm \epsfbox{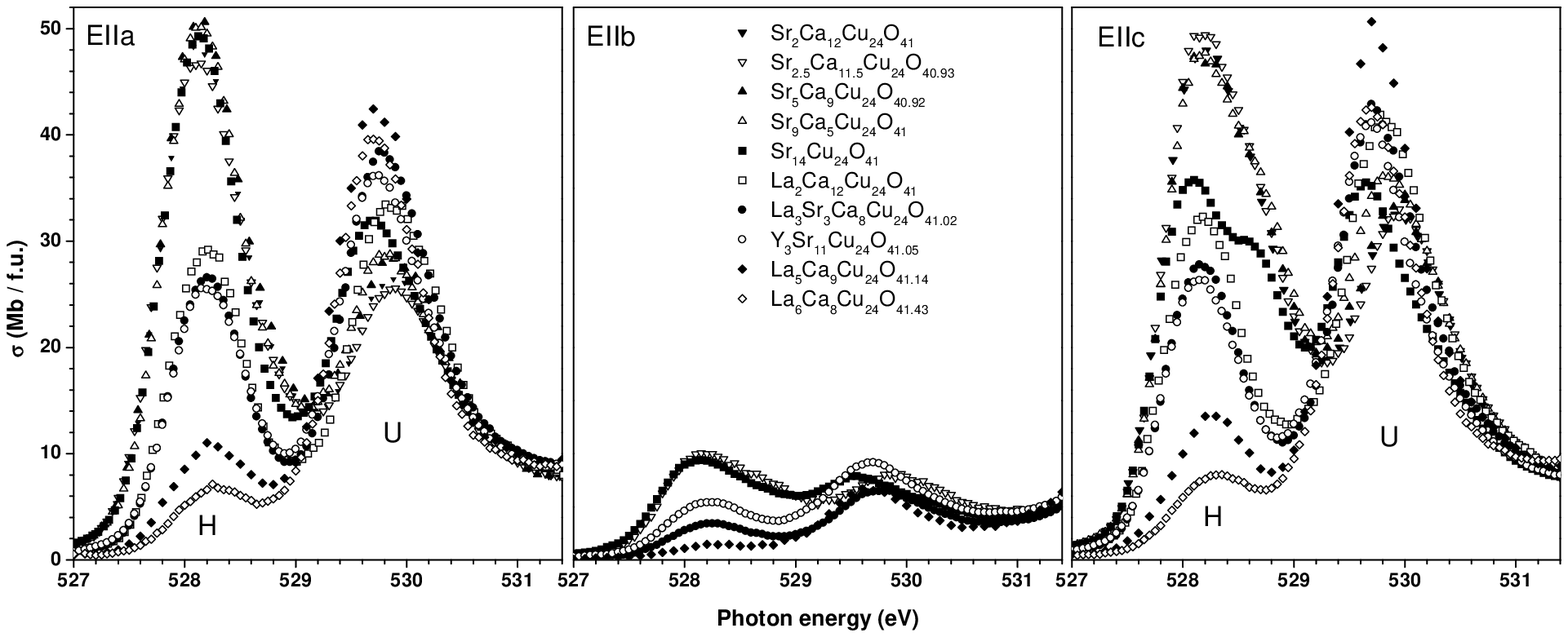}
\caption{Polarization-dependent O $1s$ NEXAFS spectra for 
La$_6$Ca$_8$Cu$_{24}$O$_{41.4}$, La$_5$Ca$_9$Cu$_{24}$O$_{41.4}$,
La$_3$Sr$_3$Ca$_8$Cu$_{24}$O$_{41}$, Y$_3$Sr$_{11}$Cu$_{24}$O$_{41}$, 
La$_2$Ca$_{12}$Cu$_{24}$O$_{41}$, Sr$_{14}$Cu$_{24}$O$_{41}$, 
Sr$_{9}$Ca$_{5}$Cu$_{24}$O$_{41}$, Sr$_{5}$Ca$_{9}$Cu$_{24}$O$_{41}$, 
and Sr$_{2}$Ca$_{12}$Cu$_{24}$O$_{41}$ for polarizations {\bf E}$\|$$a$, 
{\bf E}$\|$$b$, and {\bf E}$\|$$c$\@. The first feature (H) represents 
holes at the O sites, and the second one (U) the upper Hubbard band.
For {\bf E}$\|$$b$, only a subset of spectra is depicted to retain
clarity.} \label{alldata}
\end{minipage} 
\end{figure}

O $1s$ NEXAFS spectra for (Sr,Ca,Y,La)$_{14}$Cu$_{24}$O$_{41}$ are shown in 
Fig.\ \ref{alldata} for all three orientations {\bf E}$\|$$a$, {\bf E}$\|$$b$, and 
{\bf E}$\|$$c$ in the energy range of interest below the main edge. Concentrating 
first on the ``planar'' {\bf E}$\|$$a$,$c$ 
spectra one observes considerable spectral weight near threshold.
The first feature above threshold, around 528 eV and called H in the spectra, 
appears near the O 1$s$ binding energy as determined by x-ray photoemission 
spectroscopy (not shown)\@. Feature H, therefore, represents hole states; its 
spectral weight is related to the number of holes per unit cell. Analogous to 
the situation for HTSC materials, a second maximum is observed in the spectra 
which represents the upper Hubbard band (UHB) and is denoted by U\@. It 
appears in the O $1s$ spectra due to the Cu $3d$ -- O $2p$ hybridization 
the strength of which is described by the hopping integral $t_{pd}$\@. The 
spectral weight of feature H is smallest for La$_6$Ca$_8$Cu$_{24}$O$_{41.43}$
and grows with the Sr and/or Ca content, i.\ e., with the number of doped 
holes. At the same time, the spectral weight of feature U is seen to decrease.
Particularly noteworthy is that the NEXAFS spectra of
Y$_3$Sr$_{11}$Cu$_{24}$O$_{41}$ and La$_3$Sr$_3$Ca$_8$Cu$_{24}$O$_{41}$, both 
doped with three holes per formula unit, are identical within the experimental 
precision despite their very different Ca content; furthermore, they show 
equivalent spectral weight in H for {\bf E}$\|$$a$ and {\bf E}$\|$$c$\@. While 
for these two compounds, moreover, feature H is almost symmet-
\twocolumn[\hsize\textwidth\columnwidth\hsize\csname
@twocolumnfalse\endcsname
]

\noindent
ric in energy 
(and nearly Gaussian-shaped), one finds for the fully doped samples 
Sr$_{14-x}$Ca$_x$Cu$_{24}$O$_{41}$ a pronounced, shoulder-like asymmetry of 
H towards higher energy. An immediate conclusion that could be drawn from 
observing that these symmetries of H are present at low doping levels but  
absent at full doping is that for low doping (up to 4 holes per f.u.)\ all holes 
reside on the chain sites: only there spectral weight for {\bf E}$\|$$a$ and {\bf E}$\|$$c$
is naturally the same as both orientations measure equal projections of the {\em same}
O $2p$ orbitals ({\it cf}.\ Fig.\ \ref{structure})\@. This is in agreement 
with previous work ({\it cf}.\ Refs.\ \onlinecite{katobvs} and \onlinecite{carter})\@. 
For {\bf E}$\|$$b$, a similar double-peak structure in the energy range shown 
is observed; the corresponding spectral weights are, however, 
smaller by an order of magnitude compared to the total in-plane weight for 
{\bf E}$\|$$a$,$c$\@. This is a nice illustration of the predominantly planar 
character of the electronic structure near $E_F$ for both the chain and the 
ladder units.

These qualitative observations can be put on a somewhat firmer footing by a 
more refined data analysis based on the following ideas:
Systematic changes in H and U upon hole doping and upon Ca substitution can
clearly be observed in Fig.\ \ref{alldata}, like the developing asymmetries in H 
described above, although individual O $2p$ contributions are not well-separated 
from each other. With the present study, on the other hand, a large number of 
spectra for different compositions is available, and a promising way to 
extract as much site-specific information as possible is to perform a 
least-square fit to {\em all} measured spectra {\em simultaneously}\@. With 
the assumption that for all compositions the individual, site-specific 
spectral distributions are identical in shape, and that only their amplitudes 
and energy positions may change with composition, one can even construct 
these spectral distributions corresponding to various O orbitals in the 
fitting procedure, and thus avoid introducing artifacts through a possibly 
imperfect choice of spectral shapes for the fitting functions. Of course, 
the spectral weight and energy shifts for each composition are also 
obtained in the same fit. The spectra for
La$_6$Ca$_8$Cu$_{24}$O$_{41.43}$ and La$_5$Ca$_9$Cu$_{24}$O$_{41.14}$ were
not included in the fitting procedure for construction of the 
spectral distributions since their strong off-stoichiometry might possibly lead to 
artifacts in the results. Their partial weight, however, was calculated as well, 
using the distributions determined from all the other spectra.

In practice one is, of course, limited in the number of spectral
distributions that can reasonably be constructed in the fitting procedure; it
must be much smaller than the number of independent spectra measured (generally 
the two in-plane spectra per sample are used)\@. With two distributions each to fit 
H and U for each polarization the resulting spectral distributions for {\bf E}$\|$$a$ 
and {\bf E}$\|$$c$ turn out to be almost identical, allowing the use of the same 
distributions for both polarizations. Unlike the situation for H, where two distributions 
are {\it required} to properly approximate the experimental curve\cite{distance} and 
its doping dependence, a decomposition of U into two parts is far less justified: 
the improvement in rms deviation is only marginal, and the variation of relative 
intensities does not exhibit much connection with doping. It is thus possible to 
determine only three spectral weights: H1, H2, and U. The energies of peaks H1 and H2 
associated with the doped holes remain constant to within 30 meV and are well within 
the limits of our energy calibration ($<100$ meV); such slight energy shifts could be 
due to changes in the O $1s$ binding energy. In the case of U, a single fitting function 
had to be used (as mentioned above) despite U being a compos-
\twocolumn[\hsize\textwidth\columnwidth\hsize\csname
@twocolumnfalse\endcsname
\begin{figure}
\noindent
\begin{minipage}{178mm}
\leavevmode
\epsfxsize=178mm \epsfbox{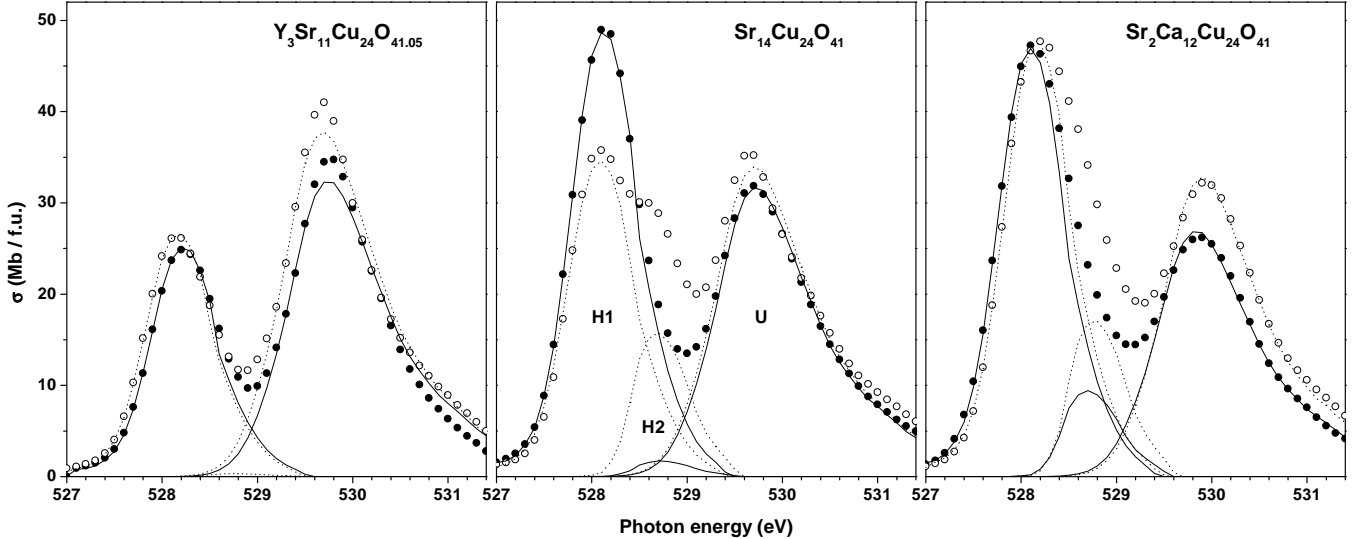}
\caption{NEXAFS spectra of Y$_3$Sr$_{11}$Cu$_{24}$O$_{41}$,
Sr$_{14}$Cu$_{24}$O$_{41}$, and Sr$_{2}$Ca$_{12}$Cu$_{24}$O$_{41}$ for 
{\bf E}$\parallel${\it a} (filled symbols) and {\bf E}$\parallel${\it c} (open
symbols) along with the fitted partial distributions corresponding to holes (H1 
and H2) and to the upper Hubbard band (U), see text. Spectral weight above 531 eV 
corresponding to states other than H and U was subtracted.\cite{footnote1}}
\label{fits}
\end{minipage} 
\end{figure}]

\noindent
ed feature, and the slightly 
larger shifts ($<200$ meV) observed for this fitting function may result from this 
simplification. The spectral weights fitted are discussed in the following.

In Fig.\ \ref{fits} NEXAFS spectra with these three fitted partial distributions 
are shown for Y$_3$Sr$_{11}$Cu$_{24}$O$_{41}$, 
Sr$_{14}$Cu$_{24}$O$_{41}$, and Sr$_{2}$Ca$_{12}$Cu$_{24}$O$_{41}$ for 
{\bf E}$\|$$a$ and {\bf E}$\|$$c$ as representative examples for the 
full data set. The partial weights extracted from such fits, corresponding 
to hole counts and the weight in the UHB, as well as their evolution
as a function of the hole count as derived from volumetry and of the Ca 
substitution level are shown in Fig.\ \ref{weight}: part (a) depicts the individual 
contributions H1 and H2 for {\bf E}$\|$$a$ and {\bf E}$\|$$c$, while part 
(b) plots the total weight of H (i.\ e., H1+H2 summed over both in-plane 
polarizations) and the analogously determined total weight for U\@. The 
left-hand side of both panels in Fig.\ \ref{weight} corresponds to increasing total
doping levels, while the right-hand side depicts fully doped (6 holes) samples 
Sr$_{14-x}$Ca$_x$Cu$_{24}$O$_{41}$\@.

Concentrating first on part (b) in Fig.\ \ref{weight}, we observe the total 
weight H to increase linearly with doping level as derived by the chemical 
analysis up to the fully
doped compound Sr$_{14}$Cu$_{24}$O$_{41}$\@. At the same time, the 
weight U of the UHB decreases. This trend in U is seen to continue
smoothly into the range of the Ca-substituted compounds -- unlike
the behavior of H, which stays constant for Ca contents greater than 4.
Turning now to the polarization-dependent decomposition of H into H1 and
H2 (Fig.\ \ref{weight} (a)), it is evident that up to and including the compound
La$_2$Ca$_{12}$Cu$_{24}$O$_{41}$ with its four doped holes, almost all the
weight is found in the lower-energy part H1, and that it is almost
the same for {\bf E}$\|$$a$  and
{\bf E}$\|$$c$ absorption. The straight line shown describes this behavior 
for H1 well. (H2 remains below 2 Mb$\cdot$eV per formula unit, and this
value may be taken as the ``threshold of significance'' for the spectral
decomposition procedure described above; the absence of significant spectral weight
other than that for H1 again illustrates that the latter represents doped holes in the 
chain.)\@ For Sr$_{14}$Cu$_{24}$O$_{41}$,
the situation changes considerably: H1 for {\bf E}$\|$$c$ lies substantially 
below its {\bf E}$\|$$a$ counterpart (which itself remains close to the 
straight line extrapolated from the lower doping levels); and
for the first time, there is a significant contribution from H2,
amounting to almost 1/3 of H1\@. The corresponding H2 for {\bf E}$\|$$a$,
on the other hand, is still almost zero. In the total weight (Fig.\ \ref{weight} (b)) the 
decrease of H1 is thus compensated. 
For the partially Ca-substituted compound Sr$_{9}$Ca$_{5}$Cu$_{24}$O$_{41}$,
H1 for {\bf E}$\|$$c$ is substantially larger than for Sr$_{14}$Cu$_{24}$O$_{41}$;
further increasing the Ca content then leaves both H1 parts almost constant, similar 
to the behavior of the total H\@. H2 for
{\bf E}$\|$$c$, on the other hand, remains constant for all Ca-substitution
levels, while H2 for {\bf E}$\|$$a$ exhibits a slight increase over the 
whole range of Ca fractions,\cite{sr5} until it reaches, for 
Sr$_2$Ca$_{12}$Cu$_{24}$O$_{41}$, about 1/6 of the value for H1.

\section{Interpretation and Discussion}

The \hfill observation \hfill that \hfill for \hfill 
La$_3$Sr$_{3}$Ca$_8$Cu$_{24}$O$_{41}$,\\  
Y$_3$Sr$_{11}$Cu$_{24}$O$_{41}$, and La$_2$Ca$_{12}$Cu$_{24}$O$_{41}$,
H consists of only one component, H1, which is furthermore almost symmetric
within the $a$$c$ plane, is readily understood if the holes at these doping 
levels almost exclusively occupy sites in the chains.
In the fully doped compound Sr$_{14}$Cu$_{24}$O$_{41}$, however, this
symmetry is obviously broken (see above)\@.
Most significant is the appearance of H2, which for {\bf E}$\|$$c$ exhibits 
considerable weight but for {\bf E}$\|$$a$ still barely exceeds the 
``threshold of significance''\@.
\begin{figure}[tb]
\noindent
\begin{minipage}[t]{86mm}
\hspace*{15mm}
\leavevmode
\epsfxsize=60mm \epsfbox{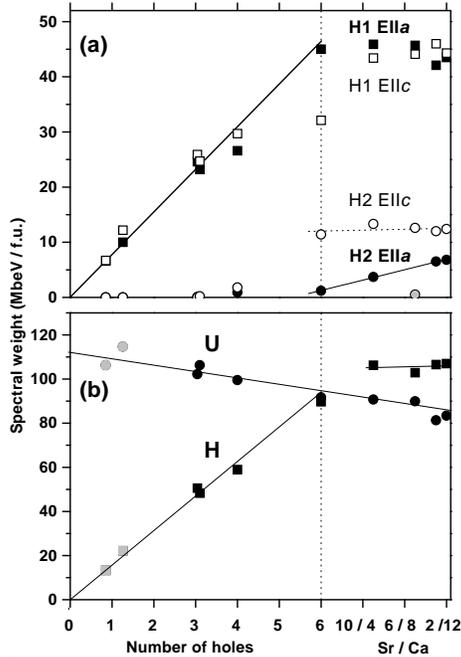}
\caption[]{Evolution of partial in-plane weights H1, H2 of holes (panel (a)),
and total spectral weight of H and U (panel (b)) for 
(Sr,Ca,Y,La)$_{14}$Cu$_{24}$O$_{41}$ compounds resulting from the least 
squares fit of three distributions to all spectra measured. Gray symbols 
correspond to data with less experimental significance, e.\ g.\ if some La 
segregation occurs (see text).\@ Lines are guides to the eye.} \label{weight}
\end{minipage} 
\end{figure}

\noindent
These observations unambiguously show that 
H2 must be spectral weight corresponding to holes on the O(1) sites of the 
legs: only there, O $2p_{\sigma}$ orbitals exist which can contribute 
strongly to {\bf E}$\|$$c$ weight while not at the same time leading to 
similarly strong {\bf E}$\|$$a$ weight, {\it cf.} Fig.\ \ref{structure}\@. 
Energetically, H2 is observed at $\approx 0.6$ eV above H1, consistent with the 
interpretation that the corresponding holes reside on O sites in different 
atomic environments. On the other hand, the {\bf E}$\|$$a$ from the O(1) leg site
should appear at an energy very similar to that for the {\bf E}$\|$$c$ contributions,
as both involve the same core level energy. While possible holes on the rung O(2) 
sites would appear for {\bf E}$\|$$a$ only it is not so clear, however, where in 
energy this contribution would show up in the spectrum. 

In the following, we present a more detailed picture of the hole 
distribution and the upper Hubbard bands, trying to account for all 
observations made so far, and check it against further experimental 
results as we proceed: 

\smallskip

(i) As discussed above, H1 is ascribed to the spectral weight of 
holes residing on the chain O(c) sites, giving a highly symmetric 
in-plane contribution in NEXAFS\@. The asymmetry between {\bf E}$\|$$a$ 
and {\bf E}$\|$$c$ at high doping levels, most prominently displayed by 
Sr$_{14}$Cu$_{24}$O$_{41}$, may have its origin in certain lattice
distortions of the chains along $c$, as observed by x-ray diffraction\cite{Cox} 
for temperatures below room temperature. These distortions appear simultaneously 
with spin dimerisation and lead to deviations from the otherwise 90$^{\circ}$ Cu-O-Cu bond 
symmetry. NEXAFS as a local probe is still sensitive to the asymmetry\cite{gap} 
induced in this way to the hole distribution of the chain, even though at
room temperature long-range order is already lost.

(ii) H2 is observed experimentally for {\bf E}$\|$$c$ and to a smaller extent 
for {\bf E}$\|$$a$ as well and is thus ascribed mainly to the spectral weight 
of holes residing on the leg O(1) sites of the ladder. 
Increasing the Ca content $x$ for the fully doped compounds 
Sr$_{14-x}$Ca$_x$Cu$_{24}$O$_{41}$ (right-hand side of Fig.\ \ref{weight} (b)) 
does not change H2 appreciably for {\bf E}$\|$$c$, indicating that 
no more holes are transferred to the O(1) $2p_z$ orbitals; the 
increase of H2 observed for {\bf E}$\|$$a$ points to a small 
but increasing number of holes that are transferred to ladder orbitals 
oriented along the $a$ axis; the data does not allow to unambiguously decide if to 
leg O(1) sites and/or rung O(2) sites. 

(iii) U consists of the spectral weight of both the chain UHB and the 
ladder UHB, and the fact that the fitting procedure described above was 
not able to separate these two contributions may indicate that the energetic 
positions and spectral distributions are too similar for all compositions. The 
in-plane Cu-O coordination and the Cu-O bond lengths in both the chain and the 
ladder units are almost identical and hence $t_{pd}$ will be to a good approximation 
the same for chains and ladders. 
In the undoped compound La$_6$Ca$_8$Cu$_{24}$O$_{41.4}$ there are 10 Cu $3d$ holes 
per f.u.\ in the chain and 14 Cu $3d$ holes per f.u.\ in the ladder unit. Hence one can 
expect the respective spectral weights in the UHB as observed in O $1s$ NEXAFS to 
be distributed accordingly: about 47:65 Mb$\cdot$eV for the chain vs.\ the ladder
UHB\@. Hole doping of the chain by about 5 holes for Sr$_{14}$Cu$_{24}$O$_{41}$ 
(see above) results in a formal Cu(c) valence of +2.5, which in most cuprates would 
result in an almost complete transfer of spectral weight from the UHB to the hole 
states H\@. 
Experimentally, we observe a reduction in the total U by $< 20$ Mb$\cdot$eV when 
going from $n_h$=0 to $n_h$=6 in the left-hand part of Fig.\ \ref{weight} (b)\@. 
Doping of just a single hole per f.u.\ into the ladder unit of Sr$_{14}$Cu$_{24}$O$_{41}$
will raise the formal Cu(l) valence only to +2.07, and the associated reduction of the 
ladder UHB will be far smaller than for the chain unit with Cu(c)$^{+2.50}$.  
We may thus assign the full reduction observed for U to the chain UHB and
find that for Sr$_{14}$Cu$_{24}$O$_{41}$, at most 45\%  
of the spectral weight for the fully developed chain 
UHB is transferred to H\@. This is considerably less than observed for 
doped CuO$_2$ planes in high-$T_c$ cuprates, and even for the closely related
doped chain structure of Sr$_{0.73}$CuO$_2$ with a formal Cu valence of 2.54 
the UHB has almost vanished.\cite{dresden5} One does find for edge-sharing chains
like the ones in the present study a Cu-O-Cu hopping that is strongly hampered as 
it is not only governed by $t_{pd}$ but is mediated by two different in-plane O 
$2p$ orbitals that are mutually orthogonal,\cite{UHB,footnote2} which
should result in a relatively slow transfer of spectral weight from the UHB to H
upon hole doping.\cite{meinders} The difference to Sr$_{0.73}$CuO$_2$ which consists 
of edge-sharing chains as well may be that there the Cu-O-Cu bond angles are in fact 
not as close to 90$^{\circ}$ as in
\begin{figure}[tb]
\noindent
\begin{minipage}[t]{86mm}
\hspace*{14mm}
\leavevmode
\epsfxsize=60mm \epsfbox{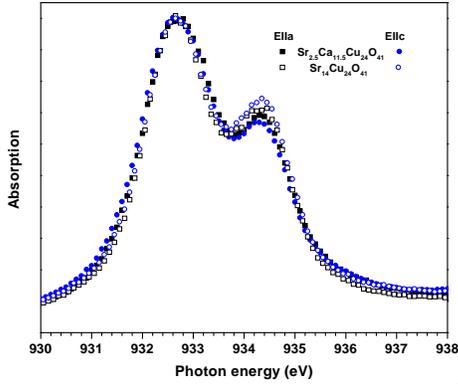}
\caption[]{Polarization-dependent Cu $2p_{3/2}$ NEXAFS for 
Sr$_{14}$Cu$_{24}$O$_{41}$ and Sr$_{2.5}$Ca$_{11.5}$Cu$_{24}$O$_{41}$, 
taken in fluorescence yield mode but with no self-absorption correction 
applied (see text)\@. All spectra are normalized to equal amplitude in the 
``white line'' around 932.5 eV\@. The ligand-hole peak at about 934 eV 
provides evidence that $t_{pd}$ is reduced for high Ca substitution for 
Sr.} \label{copper}
\end{minipage} 
\end{figure}
 
\noindent
the chains of the Sr$_{14}$Cu$_{24}$O$_{41}$ family. 

(iv) The spectral weight for {\bf E}$\|$$b$ is too small to be 
interpreted individually, especially against the backdrop of the large
contributions for {\bf E}$\|$$a$,$c$\@. Thus, even small factors like 
the only 95\% degree of linear polarization or possible small misalignments,
both mixing small fractions of the planar spectral weight to the 
{\bf E}$\|$$b$ spectra, may affect the results for {\bf E}$\|$$b$ and 
their relationship to the hole counts for the ladder sites which are 
also small. It may still be allowed to note that the spectral weight in the 
hole peak for {\bf E}$\|$$b$ tracks, in general, the nominal hole doping 
level; yet one also notes that for Sr$_{14}$Cu$_{24}$O$_{41}$ and 
Sr$_2$Ca$_{12}$Cu$_{24}$O$_{41}$, the {\bf E}$\|$$b$ component is unusually high.

(v) The observation made in the previous section that the total spectral 
weight H, which in turn is dominated by the chain H1, is significantly
larger for the Ca-substituted compound Sr$_{9}$Ca$_{5}$Cu$_{24}$O$_{41}$
than for Sr$_{14}$Cu$_{24}$O$_{41}$ is highly intriguing: stoichiometry 
requires the total number of holes to be the {\em same}! A possible solution 
for this puzzle is suggested by looking at the Cu $2p_{3/2}$ NEXAFS spectra 
for Sr$_{14}$Cu$_{24}$O$_{41}$ and Sr$_{2.5}$Ca$_{11.5}$Cu$_{24}$O$_{41}$
shown in Fig.\ \ref{copper}\@. They are recorded in fluorescence yield mode but as
self-absorption effects are exceedingly strong at this edge, corrections
like the ones applied to the O $1s$ edge (see above) are problematic 
here and thus were not performed. Total electron yield spectra, on the 
other hand, would not be reliable as the sample surfaces could not be 
cleaved {\it in situ}\@. As the large Hubbard $U$ enforces a Cu $3d^9$ 
ground state with one intrinsic hole per Cu site, it is still possible to
at least qualitatively compare different samples by normalizing the 
amplitudes of the excitonic peak (or ``white line'') around 932.5 eV\@. 
Observing a significantly {\em lower} ligand-hole peak at about 934 eV for 
the highly Ca-substituted compound -- while the corresponding weight H1 for 
O $1s$ NEXAFS is clearly {\em higher} -- may thus be interpreted as a signature 
for a reduced $t_{pd}$ hopping parameter in the chains, with the immediate 
consequence that the doped holes have increased O $2p$ character for 
heavy Ca substitution. This apparent decrease in the chain $t_{pd}$ is 
likely to be induced by strong distortions of the chain\cite{siegrist} 
on incorporation of much smaller Ca atoms, leading to highly inequivalent O 
sites in the chain with Cu(c)-O(c) distances now varying from 1.71 {\AA} to 
2.02 {\AA},\cite{siegrist} and presumably to a quite inhomogeneous 
redistribution of holes. Thus the observation of a reduced $t_{pd}$ in
NEXAFS does not stand in contrast to the structural observation of a reduced 
average Cu(c)-O(c) bond length but may be taken as an indication that the 
doped holes in the chain may be located preferably on O sites associated 
with a smaller effective $t_{pd}$.  

(vi) The fact that the spectral weight of U for the fully doped compounds
Sr$_{14-x}$Ca$_x$Cu$_{24}$O$_{41}$ seems to decrease with increasing 
$x$ is qualitatively consistent with the slight increase observed for H2 
in this substitutional regime: if holes are redistributed from chain sites to
ladder sites, the ``slow'' transfer of spectral weight between H and the UHB for the 
chains (see (iii) above) would lead to a concurrent increase in the chain UHB that 
is smaller than and thus overcompensated by the decrease in the ladder UHB\@. The 
slope of U appears larger than that of H2 and thus at first glance suggests possible 
inconsistencies. It should be noted, however, that upon Ca substitution the chain 
$t_{pd}$ is reduced (see (v) above) and lattice distortions increase,
presumably resulting in deviations from the 90$^{\circ}$ Cu-O-Cu bond geometry. 
Both effects directly lead to a reduction of the spectral weight of the chain 
UHB in O $1s$ NEXAFS and help to explain the decrease in U observed
in the right-hand part of Fig.\ \ref{weight} (b).

\smallskip

\begin{figure}[tb]
\noindent
\begin{minipage}[t]{86mm}
\hspace*{14mm}
\leavevmode
\epsfxsize=60mm \epsfbox{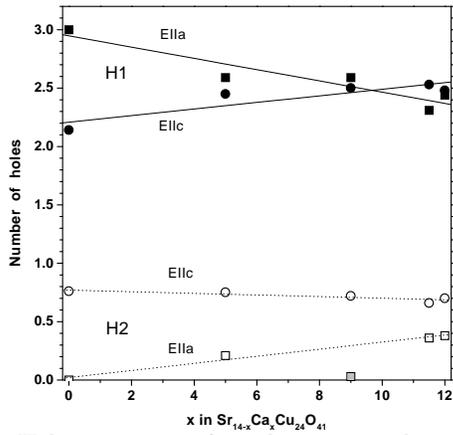}
\caption[]{Hole count per formula unit as derived from re-normalized spectral
weights in NEXAFS (see text) for Sr$_{14-x}$Ca$_x$Cu$_{24}$O$_{41}$\@. 
Shown are contributions from O $2p$ orbitals oriented along the $a$ and the 
$c$ axis in the chain and the ladder units.} \label{holes}
\end{minipage} 
\end{figure}

It is interesting to estimate the absolute hole counts $n_{i,\bf E}$ 
related to the various O orbitals, where $i$ denotes spectral weight 
corresponding to H1 or H2, and {\bf E} denotes the polarization. For each 
sample, the total spectral 
weight of H, summed over the in-plane polarizations, is directly equated to the 
corresponding total hole count as determined by the precision chemical 
analysis described above. The individual contributions 
$n_{i,\bf E}$ directly follow from this ``normalization''\@.
In this way, the problems associated to possible changes in parameters 
like $t_{pd}$, for which an indication has been presented in (v) above, 
are reduced in their effect on the individual hole occupation numbers 
determined. The results are shown in Fig.\ \ref{holes}.
The hole count most obviously associated with the ladders,
$n_{2,a}$+$n_{2,c}$, increases slightly from 0.8 holes for  
Sr$_{14}$Cu$_{24}$O$_{41}$ to 1.1 holes for
Sr$_2$Ca$_{12}$Cu$_{24}$O$_{41}$\@; considering the individual
contributions one observes that the increase is due to $n_{2,a}$ only while
$n_{2,c}$ stays virtually constant, in effect rendering the hole
distribution in ladders more two-dimensional for the heavily 
Ca-substituted compounds.

These values for fully doped compounds with high Ca content remain considerably 
smaller than the up to 2.8 holes found in optical measurements.\cite{Uchida}
The interpretation of the optical data rests on assuming zero mobility

\noindent
for 
holes on the chains, allowing one to distinguish between holes on
the chains (localized) and holes on the ladders (mobile)\@. This may be less 
valid if heavy Ca substitution leads to mobility in the chain as well, 
and the hole count for the ladders might thus be overestimated to some extent; 
but even in this case the difference between the optical and NEXAFS results 
appears too large. A point raised in Ref.\ \onlinecite{ruzicka} concerns the 
role of an effective mass greater than one which would reduce the apparent ladder
hole count deduced from the optical data; however, the assumption of 
$n_{\small chain}$=5 and $n_{\small ladder}$=1 made in Ref.\ \onlinecite{Uchida} 
to ``normalize'' their data already takes care of this. On the other hand, 
NEXAFS suffers to some extent from changes in parameters like $t_{pd}$ as pointed out 
above. A further effect could be that the holes on the rung O(2) sites might 
appear in NEXAFS at about the same energy as holes on the chain O(c) sites and thus
would, in the evaluation used here, be counted not as a ladder contribution but instead
as a chain contribution. If present at all, however, this effect is also 
expected to be small.
The differing results of optical and x-ray determinations of the hole count 
are thus noted but remain unexplained at this time.

A further remark should be made on the fact that in the original work reporting 
on superconductivity in these ladder compounds,\cite{uehara} the very compound
studied (Sr$_{0.4}$Ca$_{13.6}$Cu$_{24}$O$_{41.84}$) exhibited an excess oxygen 
content $\delta$ amounting to as much as 0.84. Although it is difficult to 
estimate the effect of up to 1.7 extra holes, it seems clear that such additional 
O atoms may lead to further substantial deformations in the crystal structure, 
and one could thus speculate that the hole concentration on the ladder sites may 
be increased considerably by the extra oxygen. Excess oxygen, on the other hand, may 
not even be necessary for superconductivity, as later reports\cite{Isobe,nagata} 
are much less specific on excess oxygen content, and we note that crystals from the
same growth rods as our Sr$_{2.5}$Ca$_{11.5}$Cu$_{24}$O$_{40.93}$
and Sr$_2$Ca$_{12}$Cu$_{24}$O$_{41}$ samples do exhibit superconductivity under high 
pressure. It may thus well be just the increased ``2D'' character of the electronic 
structure of the ladders, detected in NEXAFS as the increasingly similar H2 weight along 
$a$ and $c$, as well as the increased number of ``pseudo-apical'' O sites 
induced by the distortions, that in the end work together to support 
superconductivity in the highly Ca-substituted compounds under high pressure.

\section{Conclusion}

Polarization-dependent NEXAFS has been performed on a large number of 
single-crystalline members of the Sr$_{14}$Cu$_{24}$O$_{41}$ family of 
layered compounds containing two-leg Cu$_2$O$_3$ ladders and CuO$_2$ 
chains. Studying different substitution levels of Y or La for Sr to 
change the total number of holes has shown in a direct and 
straightforward manner that for low and intermediate hole counts, 
all holes reside on the chain sites, and that this is independent of whether 
a further fraction of Sr is replaced by isoelectronic Ca as well. For 
the fully doped compound Sr$_{14}$Cu$_{24}$O$_{41}$, on the other hand, 
approximately one hole per formula unit is observed on the ladder unit, 
specifically in O $2p$ orbitals oriented along the $c$ axis on the legs. 
Studying the fully doped substitution series 
Sr$_{14-x}$Ca$_x$Cu$_{24}$O$_{41}$, we find the total hole count in the 
ladders to increase only marginally; however, a larger fraction of O 
$2p$ orbitals oriented along the $a$ axis (on the legs and possibly on the rungs 
of the ladder) is now contributing to the spectral weight. 

\subsection*{Acknowledgements}

We greatly appreciate generous and valuable experimental help by J.-H. Park, S. 
L. Hulbert, P. Schweiss, T. Ivanova, E. Sohmen, D.-H. Lu and C. S. Gopinath. For 
illuminating discussions we are grateful to S.-L. Drechs\-ler and M. Knupfer. 
Research was carried out in part at the National Synchrotron Light Source, 
Brookhaven National Laboratory, which is supported by the U. S. Department of 
Energy, Division of Material Sciences and Division of Chemical Sciences, under
contract number DE-AC02-98CH10886.
V. C. was supported by the Office of Naval Research.


\begin{references}
\bibitem{distance} The energy shift between H1 and H2 was treated as a fit parameter as
 well but was assumed to be the same for all spectra.
\bibitem{footnote1} To obtain spectra for the fitting procedure that consist solely 
 of H and U, the higher-energy part of each spectrum was removed by approximating it, 
 for the region of U and up to 534 eV, by a Gaussian which was then subtracted.
\bibitem{footnote2} In Ref.\ \onlinecite{neudert}, transfer integrals in the range 30
 $\dots$ 80 meV along the chains of Li$_2$CuO$_2$ are presented, which are smaller by
 a factor of at least 7 compared to the value (0.55 eV) for the corner-sharing chain
 compound Sr$_2$CuO$_3$ where Cu-O-Cu hopping is not hampered.
\end{references}
\end{document}